\documentstyle[aps, graphicx]{revtex}

\begin{document}

\title{Motion of dark solitons in trapped Bose-Einstein condensates}

\author{Th. Busch and J. R. Anglin}

\address{Institut f\"ur Theoretische Physik, 
         Universit\"at Innsbruck, 
         A--6020 Innsbruck, AUSTRIA}

\maketitle

\begin{abstract}  
  We use a multiple time scale boundary layer theory to derive the
  equation of motion for a dark (or `grey') soliton propagating
  through an effectively one-dimensional cloud of Bose-Einstein
  condensate, assuming only that the background density and velocity
  vary slowly on the soliton scale.  We show that solitons can exhibit
  viscous or radiative acceleration (anti-damping), which we estimate
  as slow but observable on experimental time scales.
\end{abstract}

\pacs{PACS number(s): 03.75.Fi, 03.65 Ge}

\date{\today}

The success of the Gross-Pitaevski mean field theory in describing
experimentally observed dilute Bose condensates \cite{Science95} shows
that one really can persuade a large number of particles to behave as
a field.  There is thus a pleasant circularity in investigating
situations where this field in turn behaves in a particle-like manner,
in that it contains topological defects or solitons.  In this paper we
discuss one particular particle-like configuration of the
Gross-Pitaevski mean field, namely the one-dimensional {\it dark
  soliton}.  Quasi-one-dimensional traps are realistic prospects in
the relatively near future \cite{wolfgang}, and dark solitons are
expected to emerge in them from generic violent collisions between
condensates \cite{Clark,Burnett}.  A controlled method for creating
them by adiabatic state engineering with lasers has also recently been
proposed \cite{ADP}.  And they are expected to play a crucial role in
the eventual decay of superfluid currents in tight toroidal traps
\cite{Ericetal}, which would be a valuable analog of the thin
superconducting wires whose resistivity is one of the triumphs of
non-equilibrium statistical mechanics \cite{Langeretal}.  Although
dark solitons have been studied extensively in nonlinear optics
\cite{DSrep}, optical fibres are spatially homogeneous on the relevant
scale.  In this Letter we extend or correct previous treatments of
dark soliton motion in Bose condensates \cite{Clark,Huo,Morgan}, by
using multiple scale analysis to derive equations of motion for a dark
soliton moving through a background condensate which changes slowly in
both space and time, and is subject to a generic slowly-varying
potential (not necessarily harmonic).  This powerful analytical method
may also be useful for other structures.

The Gross-Pitaevski equation (GPE) governs the evolution of the
c-number `macroscopic wave function' $\psi(\vec{x},t)$ of a
Bose-Einstein condensate.  (This is of course a mean field
approximation to the full quantum field theory; we will consider
dissipation from quasi-particle interactions very briefly below.)
Incorporating a chemical potential by extracting a factor $e^{-i\mu
  t/\hbar}$, and then appropriately scaling the wave function, space,
and time, one can write this equation in the convenient form
\begin{equation}
 \label{GPE} 
 i\partial_t\psi = -{1\over2}\nabla^2\psi 
                   +(|\psi|^2 +V(\vec{x})-1)\psi\;.  
\end{equation} 
We assume here a positive scattering length; and we do not restrict
the normalization constant $U\equiv\int\!dx\,|\psi|^2$, which is the
number of particles rescaled by the strength of their mutual
repulsion.  Crucially, we assume a trap so thin that one can apply the
GPE in one dimension.  The approach to this limit from three
dimensions has recently been discussed \cite{Pethick,Shlyapnikov}.
The essential requirement is that the transverse thickness of the trap
be less than the healing length, to stabilize against buckling modes
in the GPE.  Making transverse confinement stronger than the
temperature will make even the quantum field theory effectively one
dimensional.  Experimental capability is already approaching both
these limits.

Eqn.~(\ref{GPE}) in one dimension with constant $V$ has been
extensively studied in nonlinear optics \cite{DSrep}, and a solution
with a localized structure has long been known: $\psi_{DS} = \tanh
\sqrt{1-V}x$.  This time-independent solution is known as a {\it dark
  soliton}, because it describes a small dark spot in a light pulse;
in our case this becomes a small `bubble' of low condensate density in
the dilute Bose gas.  If $\psi(x)$ were restricted to be real, the
dark soliton would be topologically stable, like other `kink'
solitons; but by taking $\psi$ into the complex plane one can deform
it into a configuration with constant density and phase, eliminating
the `bubble'.  So unlike two-dimensional vortices, dark solitons are
not topologically stable.  Before considering their motion, therefore,
one should first examine their stability; but in fact the two problems
are closely connected, because the complex deformations of $\psi_{DS}$
include the larger family of dark solitons moving with arbitrary
(sub-critical) velocities \cite{vortexnote}.  These exact
non-stationary solutions to (\ref{GPE}), moving with constant velocity
$p\equiv \dot{q}$, are
\begin{equation}
 \label{grey}
 \psi_{GS} = ip + \sqrt{v_c^2-p^2}\tanh\sqrt{v_c^2-p^2}[x-q(t)]\;, 
\end{equation}
where $v_c\equiv\sqrt{1-V}$ is the Landau critical velocity (which the
soliton cannot exceed).  For $p\to 0$ we recover the motionless dark
soliton at position $x=q$.  Since for non-zero $p$ the condensate
density $|\psi|^2$ never vanishes, moving dark solitons are also
called {\it grey solitons}.

For moving solitons the difference between maximum and minimum
densities is $v_c^2-p^2$, and the phase slip across the soliton is
$\pi +2\arctan (p/\sqrt{v_c^2-p^2})$.  This means that in the limit
$p\to\pm v_c$, the soliton becomes identical with motionless
condensate.  Thus the soliton with maximum speed is the ground state;
the energy of slower solitons is higher!  In this sense one may say
that {\it dark solitons have negative kinetic energy.}  To be precise:
The solutions given in (\ref{grey}) have fixed chemical potential (set
to one), and their free energy $G\equiv E-U$ is
\begin{equation}
 \label{Ggrey} 
  G = G_0 +{4\over3}(1-V-p^2)^{3\over2}\;,  
\end{equation} 
where $E \equiv {1\over2}\int\!dx\,[|\psi'|^2 + |\psi|^4 +
2V|\psi|^2]$, and $G_0$ is the free energy of the ground state $\psi =
\sqrt{1-V}$.  (Alternatively we could change the chemical potential
with $p$ so as to keep the particle number constant; this slightly
different family of grey soliton solutions with constant $U$ has
energy $E=E_0+(4/3)(1-V-p^2)^{3/2}$.)  Thus, dark solitons are
energetically as well as topologically unstable; but their instability
is to acceleration, not to filling or collapse.  Bogoliubov theory
shows that acceleration is indeed their only instability, and as we
discuss below, the `anti-damping' time scale should be quite long.

So apart from slow anti-damping, dark solitons in bulk behave as
robust free particles, obeying $\ddot{q}=\dot{p}=0$.  We now consider
a dark soliton in a slowly varying medium, where we will be able to
derive a more complicated equation of motion, if we interpret the
`slow variation' of $V(x)$ as implying that there exists a length
scale $\Lambda$ which is both large compared to the soliton scale and
small compared to the trap scale.  Precisely: there is a small
dimensionless $\epsilon$, such that $\exp(-\sqrt{1-V(q)-p^2} \Lambda)
<< \epsilon$ for all phase space points $(q,p)$ through which the
soliton will actually pass, but $V(x) = V(q) + V'(q) (x-q)+ {\cal
  O}(\epsilon^2)$ as long as $|x-q| < \Lambda$ (with $V'(q)|\Lambda|$
being of order $\epsilon$).  We will then examine an interval
$|x-q|<\Lambda$ around a grey soliton in a trap, the interval moving
with the soliton, and smoothly patch this interval into a background
condensate cloud in the hydrodynamic limit.  Applying a simple form of
multiple time scale analysis will then yield the equation of motion.
This involved procedure (`boundary layer theory') is indeed necessary:
merely treating $V(x)$ as a perturbation is only valid if the
potential is everywhere small, whereas we are interested in cases
where, over large enough distances, it can change greatly.  And
ordinary perturbation theory will be valid only for a short time, but
we are interested in large changes over longer times (such as the
reflection of the soliton from a barrier).

We begin with the simplest step of considering the background cloud.
We will assume that the background cloud consists of condensate
varying slowly on the healing length scale and its associated time
scale, except possibly for small high frequency perturbations.  For
the dominant low-frequency component, we define $\psi =
\sqrt{\rho}e^{i\theta}$ for real $\rho,\theta$, and stipulate that
spatial and temporal derivatives of $\rho$ and
$v\equiv\partial_x\theta$ are of order $\epsilon$.  We may therefore
neglect $\sqrt{\rho}''/\sqrt{\rho}$ in the GPE to obtain the
hydrodynamic equations
\begin{eqnarray}
 \label{hydro}
 \partial_{t} \rho &=&     - \partial_{x} (\rho\partial_{x} \theta)  
                   \equiv  -\partial_x (\rho v)\nonumber\\
 \partial_t \theta &=&     1- \rho -V - v^2/2\;.
\end{eqnarray}

We now patch our family of solitons into this background condensate:
within $|x-q|<\Lambda$ we write $\psi = e^{i\bar\theta +
  i\bar{v}(x-q)}[\psi_0 + \epsilon \psi_1(x-q,t)] + {\cal
  O}(\epsilon^2)$, for
\begin{equation}
 \psi_0 = i(p-\bar{v}) + \kappa\tanh\kappa(x-q)\;,
\end{equation}
where $\bar{\theta}(t)\equiv
[\theta(q-\Lambda,t)+\theta(q+\Lambda,t)]/2$, $\bar{\rho}$ and $\bar
v$ are similarly defined, and $\kappa^2 \equiv \bar\rho - (p-\bar
v)^2$.  Since $p$ would be constant if $\epsilon\to 0$, we conclude
that $\dot{p}$ is order $\epsilon$; in fact, $p$, $\bar v$, and
$\kappa$ may be taken as functions of the `slow time' $\epsilon t$.

We then expand the Gross-Pitaevski equation to order $\epsilon$ within
$|x-q|<\Lambda$, keeping in mind that $\dot{p}, V', \dot\kappa$ etc.,
are all ${\cal O}(\epsilon)$.  First using (\ref{hydro}) to establish
\begin{eqnarray}
 \label{thetadot}
 \partial_t\bar{\theta} &=& {1\over2}\sum_\pm
  \bigl[(\partial_t + p\partial_x)\theta(x,t)\bigr]_{x=q\pm\Lambda}\nonumber\\
  &=& p \bar{v} + 1-\bar{\rho}-V(q) - \bar{v}^2/2 + {\cal O}(\epsilon^2)\;,  
\end{eqnarray}
we find from the zeroth order terms $\dot{q} = p$ as always, plus the
following at first order in $\epsilon$:
\begin{eqnarray}
 \label{tadaaa}
 &&[ V'(q) + \dot{\bar v}](x-q)\psi_0
 +[\dot{p}-\dot{\bar v} - i\dot{\kappa}{\partial\ \over\partial\kappa}
 \kappa\tanh \kappa(x-q)]\nonumber\\
 &&=\epsilon\bigl[i\partial_t{\psi}_1\vert_{x-q} 
                  -i(p-\bar{v})\psi'_1+ {1\over2}\psi_1'' - 
                  \psi_0^2\psi_1^*-(2|\psi_0|^2 -\bar{\rho})\psi_1\bigr]\;.
\end{eqnarray}
We will abbreviate Eqn.~(\ref{tadaaa}) as
$i\partial_t{\psi_1}\vert_{x-q} + {\cal E}(\psi_1,\psi_1^*) = {\cal
  J}(x,\epsilon t)$.  (Note that it is a straightforward but very
important step in obtaining (\ref{tadaaa}) to distinguish
$\partial_t$, which is, as usual, differentiation with respect to $t$
with $x$ fixed, from differentiation with respect to $t$ with $x-q$
fixed: $\partial_t f(x-q,t)\vert_{x} = \partial_t f(x-q,t)\vert_{x-q}
- p\partial_x f(x-q,t) \vert_t$ for any function $f$.)

We could then proceed to solve Eqn.~(\ref{tadaaa}) using the Green's
function for the homogeneous part.  To construct this we would need
all the independent solutions to the homogeneous equation; but in fact
for our purpose we will require only the four independent solutions
$u_1,...,u_4$ to the time-independent equation ${\cal
  E}(u_j,u_j^*)=0$.  Distinguishing the fast and slow parts of
$\psi_1$ by defining $\psi_1 \equiv \phi(x-q,\epsilon t) +
\chi(x-q,t)$, we can use (\ref{tadaaa}) to show that the real parts of
certain integrals are constrained to vanish:
\begin{eqnarray}\label{slow}
&& {\rm Re} \int_{q-\Lambda}^{q+\Lambda}\!dx\, 
\Bigl(2iu_j^*(x)\partial_t\chi\vert_{x-q} + \partial_x[\chi' u_j^* - u_j^{*'}\chi - 2i(p-\bar{v})
u_j^*\chi]\Bigr) 
\nonumber\\
&=&{\rm Re}\int_{q-\Lambda}^{q+\Lambda}\!dx\,\Bigl(2u_j^*(x) {\cal J}(x,\epsilon t) 
-\partial_x[\phi' u_j^* - u_j^{*'}\phi - 2i(p-\bar{v})u_j^*\phi]\Bigr)=0\;.
\end{eqnarray}
Here the crucial final equality follows from the fact that the two
sides of the preceding equation vary on different time scales, and so
must separately equal zero.  (Since the first line is linear in
$\chi$, which must be fast, a non-zero constant is not allowed.)  This
is the great strength of the combination of boundary layer and
multiple time scale analysis, that it allows us to obtain the motion
of a short-scale defect in a long-scale background, by solving only
time-independent equations.

Eqn.~(\ref{slow}) gives us four constraints, which since all four
$u_j(x)$ may be obtained explicitly, can be evaluated.  In addition we
require that our soliton $\psi$ match smoothly into the background
flow as $|x-q|\to\Lambda$, and this introduces constraints from
(\ref{hydro}) as well.  Together these constraints fix the hitherto
unknown $\dot{p}$, and also relate $\rho, \theta, v$ at $x=q-\Lambda$
to their values at $x=q+\Lambda$.  We illustrate the procedure with
the simplest but most important constraint, the one involving $u_1(x)
= \,{\rm sech}^2\,\kappa(x-q)$.  Since $u_1(\pm \Lambda)$ and
$u'_1(\pm \Lambda)$ are exponentially negligible, we discard terms of
this order in (\ref{slow}).  We can then extend the limits of
integration to infinity and shift the integration dummy variable $x-q
\to y$, to obtain
\begin{eqnarray}\label{IT}
\kappa\int_{-\infty}^\infty\!dy\,[\kappa(V'(q)+\dot{\bar
  v})y\tanh\kappa y +\dot{p}-\dot{\bar v}]\,{\rm sech}^2\kappa y 
 = 2\dot{p} + V'(q) - \dot{\bar v} = 0\;.
\end{eqnarray} 
This is the equation of motion, accurate to ${\cal O}(\epsilon)$, for
a dark soliton in an otherwise hydrodynamic condensate in an
inhomogeneous potential.  We will examine it in some simple limits,
before discussing the conditions obtained from the other $u_j$, and
from requiring (\ref{hydro}) as $|x-q|\to\Lambda$.

\begin{figure} 
 \centering
 \includegraphics[clip=false,width=0.6\linewidth]{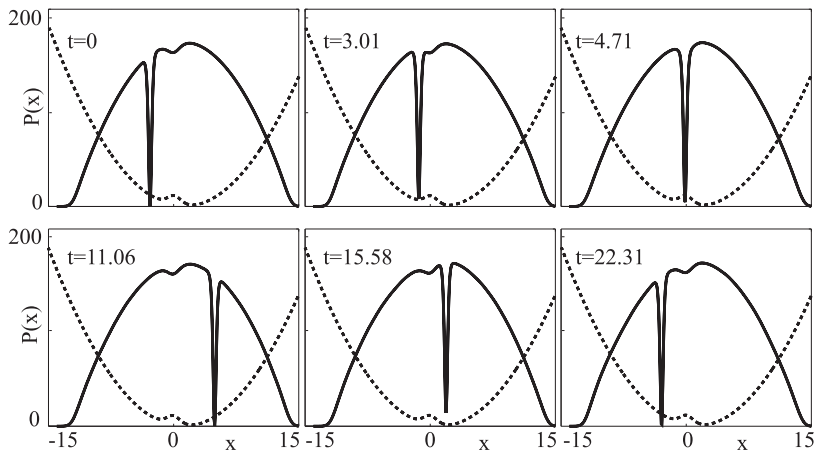}
 \caption{Density $|\psi|^2$ for a dark
   soliton oscillating through a static Thomas-Fermi cloud at $\int dx
   |\psi|^2=300$, with potential $V=0.1 x(x-2)+1.1\,{\rm sech}^2x$
   shown in dots.  Initially $ q=-2.18$.  Eqn.~(\ref{EofM}) thus
   predicts the time between turning points (where $|\psi|^2=p^2=0$ at
   the minimum) to be $T/2=11.5$; the error of about 4\% is indeed
   $\cal{O}(\epsilon)$.}
 \label{fig:SolOsc}
\end{figure}

With $v=0$, Eqn.~(\ref{IT}) implies 
\begin{equation}\label{EofM}
\ddot{q} = - {1\over2}V'(q)\;.
\end{equation}
In a harmonic trap, this implies oscillation of the soliton with
frequency $1/\sqrt{2}$ times that of the dipole mode of the condensate
(the trap frequency)\cite{Clarkcom}.  This result can also be obtained
for small oscillations by solving the Bogoliubov equations for a
motionless soliton in a trap, using a simpler, time-independent
version of the `boundary layer' approach that led to (\ref{IT})
\cite{Shlyapnikov}.  We have confirmed this frequency to rather more
than the expected accuracy in numerical simulations \cite{numerical}
of harmonic traps over a wide range of condensate densities and
oscillation amplitudes; we have also confirmed that the center of mass
is decoupled and oscillates at the trap frequency.  Eqn.~(\ref{EofM})
also holds for arbitrary potentials, however, as long as they vary
slowly on the healing length scale.  We have therefore further
confirmed the good accuracy of our equation of motion by solving
Eqn.~(\ref{GPE}) numerically over a wide range of parameters and for
various potentials; a generic example is shown in
Fig.~\ref{fig:SolOsc}.  Since with lasers one can generate micro-wells
or barriers in a trap, it should be possible to realize similar
potentials experimentally.

We now consider a stationary background flow, such as in an
inhomogeneous toroidal trap holding a persistent current.  In general
the system is quite complicated; but in the limit where both the
inhomogeneous potential $V$ and the average kinetic energy $v_0^2$ are
small compared to the chemical potential, we have $\rho\doteq 1-V$, $v
\doteq v_0 [1+V]$, which with $\partial_t v=0$ implies the easily
solvable equation
\begin{equation}\label{weird}
\ddot{q} = V'(q)[v_0 \dot{q}-1]/2\;.
\end{equation}
Despite the $\dot q$ term, Eqn.~(\ref{weird}) is not dissipative: it
may be derived variationally from the Lagrangian
$(2/v_0^2)(1-v_0\dot{q})[\ln(1-v_0\dot{q})-1] - V$, and the energy
$\dot{q}{\partial L\over\partial\dot{q}} - L$ is conserved.

A simple example of the generally still more complex case where $\rho$
and $v$ are time-dependent is a soliton moving in a harmonic trap of
frequency $\Omega$ in which the collective dipole mode has also been
excited:
\begin{equation}\label{dipole}
\ddot{q} = -{\Omega^2\over 2}[q+Q\cos\Omega(t-t_0)]\;,
\end{equation}
where $Q$ is the dipole amplitude.  As required by the Ehrenfest
theorem for a condensate in a harmonic trap, the rigid dipole
oscillation of background and soliton together,
$q=Q\cos\Omega(t-t_0)$, is a solution to (\ref{dipole}).

Since this Ehrenfest theorem states that the centre of mass of the
condensate must oscillate at the trap frequency $\Omega$, but
(\ref{EofM}) makes the small soliton `bubble' oscillate at
$\Omega/\sqrt{2}$, it is clear that the background condensate must be
perturbed by the soliton moving through it.  This brings us back to
the constraints we have not yet examined, which turn out to imply
discontinuities of ${\cal O}(\epsilon)$ in both $\rho$ and $v$ between
$x-q =\pm\Lambda$.  These are in addition to the trivial
discontinuities $\propto \Lambda$ due to background gradients.  It is
both convenient, and consistent with our ${\cal O}(\epsilon)$,
`boundary layer' approach, to consider the entire interval
$|x-q|<\Lambda$ to be pointlike as far as the background condensate is
concerned; so, formally letting $\Lambda \to 0$ after obtaining all
our results so far, the discontinuities across the soliton become
abrupt.  The requirement for them can then be expressed as delta
function sources, at $x=q(t)$, which must be added to the hydrodynamic
equations.  The result can be shown to be
\begin{eqnarray}\label{rxn}
\partial_t \rho &=& - \partial_x(\rho v) + 2\dot\kappa \delta(x-q)\nonumber\\
\partial_t v &=& -\partial_x(v^2/2 + \rho + V) + \rho^{-1}  
\delta(x-q)[\kappa (V' + \dot{v})
+ 2(p-v)\dot\kappa]\;.
\end{eqnarray}
In most cases indeed these delta function sources are unimportant,
since the soliton couples only to the smooth part $\bar{v} =
[v(q+,t)+v(q-,t)]/2$, and the sources generate only discontinuities.
The effect of these on $\bar v$ depends on the boundary conditions for
the entire condensate, and solving (\ref{IT}) and (\ref{rxn}) together
to determine this effect is generally not much easier than numerically
solving the GPE with the dark soliton.  There are nevertheless some
important points that can be learned from the source terms.  For
instance, they preserve the Ehrenfest theorem in a harmonic trap, as
may be checked straightforwardly by evolving $X = -2\kappa q +
\int\!dx\, x\rho$ under (\ref{IT}) and (\ref{rxn}).  And because of
its coupling to the background fluid, one can deduce that a dark
soliton oscillating in a small well within a large sample of bulk
condensate will generate sound waves, and so exhibit radiative
anti-damping.  Numerical integration of the GPE confirms this
prediction: the soliton eventually escapes from the micro-well, the
radiation ceasing as it enters the region of constant potential
\cite{BGP}.

In a finite trap, however, coupling to the background condensate modes
does not provide dissipation.  In this case dissipation can only come
from corrections to mean field theory; in particular, from collisions
with uncondensed atoms of the thermal cloud.  A simple estimate of the
anti-damping time scale is provided by the rate at which the soliton
encounters particles, divided by the number $2\kappa$ of particles
`in' the soliton (for the `soliton mass').  At current experimental
temperatures and densities, with 99\% of the particles in the
condensate, this time is on the order of one second; which agrees with
the calculation in Ref.~\cite{Shlyapnikov} of the dark soliton decay
time.  It is clear therefore that the instability of dark solitons is
by no means fast enough to prevent their observation.

\begin{center}{\bf Acknowledgements}\end{center}

We are happy to acknowledge valuable discussions with J.I.~Cirac,
V.~Perez-Garcia, and P.~ Zoller.  This work was supported by the
European Union under the TMR Network ERBFMRX-CT96-0002 and by the
Austrian FWF.






\begin{references}


\bibitem{Science95}
M.H.~Anderson, J.R.~Ensher, M.R.~Matthews, C.E.~Wieman and E.A.~Cornell,
Science, {\bf 269}, 198 (1995);
C.C.~Bradley, C.A.~Sackett, J.J.~Tollett and R.G.~Hulet, Phys. Rev. Lett. {\bf 75}, 1687 (1995);
K.B.~Davis, M.-O.~Mewes, M.R.~Andrews, N.J.~van~Druten, D.S.~Durfee, 
D.M.~Kurn and W.~Ketterle, Phys. Rev. Lett. {\bf 75}, 3969 (1995). 

\bibitem{wolfgang} W.~Ketterle and A.~Aspect, private communications.

\bibitem{Clark} 
W.P.~Reinhardt and C.W.~Clark, J. Phys. B {\bf 30}, L785 (1997).

\bibitem{Burnett}
T.F.~Scott, R.J.~Ballagh and K.~Burnett, J. Phys. B {\bf 31}, L329-335 (1998).

\bibitem{ADP}  R.~Dum, J.I.~Cirac, M.~Lewenstein, and P.~Zoller, Phys. 
Rev. Lett. {\bf 80}, 2972 (1998).

\bibitem{Ericetal} E.J.~Mueller, P.M.~Goldbart and Y.~Lyanda-Geller,
  Phys. Rev. {\bf A57}, R1505 (1998).

\bibitem{Langeretal} J.~Langer and V.~Ambegaokar, Phys. Rev. {\bf
    164}, 498 (1967);  D.~McCumber and B.~Halperin, Phys. Rev. {\bf
    B1}, 1054 (1970).

\bibitem{DSrep} Y.S.~Kivshar and B.~Luther-Davies, Phys. Rep. {\bf 298}, 81 
(1998). 

\bibitem{Huo}
T.~Hong, Y.Z.~Wang and Y.~S.~Huo, Phys. Rev. A  {\bf 58}, 3128 (1998). 

\bibitem{Morgan}
S.A.~Morgan, R.J.~Ballagh and K.~Burnett, Phys. Rev. {\bf A55}, 4338 (1997).

\bibitem{Pethick}
A.D.~Jackson, G.M.~Kavoulakis and C.J.~Pethick, Phys. Rev. A  {\bf 58}, 2417 
(1998). 

\bibitem{Shlyapnikov} A.E. Muryshev, H.B. v. Linden v.d. Heuvell and G.V. Shlyapnikov,
cond-mat/9811408; P.O. Fedichev, A.E. Muryshev and G.V. Shlyapnikov, 
cond-mat/9905062.

\bibitem{vortexnote}  In contrast, the
velocity of vortices relative to the ambient superfluid is fixed by
the local gradient of the background density: the
price of topological stability is a reduced phase space, in which
vortex $x$ and $y$ co-ordinates are canonically conjugate to each
other.  See B.Y.~Rubinstein and L.M.~Pismen, Physica D {\bf 78}, 1 (1994).
 
\bibitem{Clarkcom} An equation similar to (\ref{EofM}) is stated
  without derivation in Ref.~\cite{Clark}, but (when translated into
  our units) without the factor of $1/2$.  This discrepancy may be
  seen in a co-ordinate-free way, by noting that in a harmonic trap
  the GPE as written in \cite{Clark} implies the same frequency for
  the collective dipole mode as is given by \cite{Clark}'s soliton
  equation of motion.  The same equation found in \cite{Clark} is derived in
  Ref.~\cite{Morgan} by assuming that the soliton does not move
  relative to the background; this may only be achieved in a locally
  harmonic trap, in which case the result  agrees with our 
  Eqn.~(\ref{dipole}).  Ref.~\cite{Huo} proposes
  $\dot{q}[\rho(q)]^{-1/2}$ constant, but mentions that oscillation
  actually occurs instead.

\bibitem{numerical}
We use the split-operator technique described in
J.A.C.~Weideman and B.M.~Herbst, SIAM J. Numer. Anal. {\bf 23}, 485 (1986).



\bibitem{BGP} Th.~Busch and J.R.~Anglin, in preparation.
\end{references}
\end{document}